\documentclass[aps,10pt,nofootinbib]{revtex4}
\usepackage{epsfig,citesort,float, verbatim}

\begin{document}

\begin{center}
{\Large \bf Folding of Cu, Zn superoxide dismutase and\\
Familial Amyotrophic Lateral Sclerosis}

\vspace*{1cm}

{Sagar D. Khare$^*$, Feng Ding$^{\dagger}$ and Nikolay V. Dokholyan$^*$\footnote{Corresponding author, Email: dokh@med.unc.edu}}
\vspace*{0.5cm}

{$^*$Department of Biochemistry and Biophysics, \\ University of
North Carolina at Chapel Hill, \\ School of Medicine, Chapel Hill, NC
27599, USA\\
$^{\dagger}$Center for Polymer Studies, Department of Physics,\\ Boston
University, Boston, MA 02215, USA}

\date{\today}
\end{center}
\smallskip

\vspace*{2cm}

\centerline{\Large\bf Summary}

{\bf Cu,Zn superoxide dismutase (SOD1) has been implicated in the familial form of the neurodegenerative disease Amyotrophic Lateral Sclerosis (ALS). It has been suggested that mutant mediated SOD1 misfolding/aggregation is an integral part of the pathology of ALS. We study the folding thermodynamics and kinetics of SOD1 using a hybrid molecular dynamics approach. We reproduce the experimentally observed SOD1 folding thermodynamics and find that the residues which contribute the most to SOD1 thermal stability are also crucial for apparent two-state folding kinetics. Surprisingly, we find that these residues are located on the surface of the protein and not in the hydrophobic core. Mutations in some of the identified residues are found in patients with the disease. We argue that the identified residues may play an important role in aggregation. To further characterize the folding of SOD1, we study the role of cysteine residues in folding and find that non-native disulfide bond formation may significantly alter SOD1 folding dynamics and aggregation propensity.} 
\medskip

\noindent {\bf Keywords:} Cu,Zn superoxide dismutase, Familial Amyotrophic Lateral Sclerosis, molecular dynamics simulations, aggregation, misfolding.
\clearpage

\section{Introduction}

ALS is the most common motor neuron disease in human adults
 that is characterized by selective motor neuron death \cite{Cleveland01, Beckman01, Rowland01, Siddique91}. In
approximately 10\% of the cases of ALS, the disease is inherited, thus
called Familial ALS (FALS) \cite{Cleveland01}.
Mutations in the cytoplasmic enzyme SOD1 were identified 
as the primary cause of approximately 20\% of FALS cases
\cite{Rosen93, Gurney94}. More than 90 mutations have been identified so far.
The FALS mutations are scattered throughout the primary sequence and 
three-dimensional structure of SOD1, which exists as a homo-dimer in
native state (each monomer is 153 amino acids in length)
\cite{Gaudette00}. One hypothesis to explain the toxic gain-of-function
of the mutants is that the toxicity is derived from intracellular
aggregates and/or failure of SOD1 degradation, which is supported by the observation that in both mice and
cell culture models, death of mutant neurons is preceded by formation
of cytoplasmic aggregates containing mutant SOD1 \cite{Johnston00, Bruijn98, Durham97}. In addition, SOD1 knockout mice do not develop motor neuron disease \cite{Reaume96}.
Aggregation or misfolding is therefore a characteristic of
SOD1 mediated FALS. Toxicity may arise through (i) aberrant chemistry
due to misfolded mutants \cite{Bromberg99,Cleveland99, Elliot01}, (ii) saturation
of essential cellular machinery such as chaperones and proteasome
components \cite{Tran99, Bruening01}, or (iii) oligomeric forms of the aggregate may themselves be toxic to the cells, as has been suggested for other
neurodegenerative diseases \cite{Walsh02}.

Aggregation of proteins is mediated by a variety of factors including
native state stability \cite{Chiti00, Canet02}, \(\beta\)-sheet propensity \cite{Ding02b}, net charge and overall hydrophobicity \cite{Chiti02a}. In addition, folding and aggregation are kinetically competing processes \cite{Chiti02b, Ding02b, Dobson99}. Thus, it is crucial to study the folding thermodynamics and kinetics of SOD1 in order to understand the causes of SOD1 misfolding, aggregation and toxicity. A microscopic picture of SOD1 folding and/or aggregation dynamics of SOD1 is not available from either theory or experiment.  We reconstruct the SOD1 folding mechanism from molecular dynamics simulations.

Neither the structure of SOD1 aggregates nor the mechanism of
aggregation is fully understood.  Several experimental studies have
characterized the folding thermodynamics of SOD1 and the associated
FALS mutants \cite{Hallewell91, McRee90, Parge92, Lepock90, Stroppolo00,
Mei92, Rodriguez02}. Lepock et al. \cite{Lepock90} and Rodriguez et
al. \cite{Rodriguez02} measured the stability of mutants in differential
scanning calorimetry studies of the irreversible unfolding of the
distinctly metallated species of the enzyme, and showed that mutants
are destabilized relative to the wild type. In addition, a
crystal structure of the G37R mutant \cite{Hart98} shows higher atomic
displacement parameters (B-factors) for the side-chains compared to 
the wild type, but the backbone conformation is not significantly
different, thus indicating greater molecular flexibility in some
portions of the structure. It was recently shown that the for some mutants, the apo-state (without metals) of the mutants is markedly destabilized compared to that of the wild-type, while the holo-state (with metals) of both mutants and wild-type is not significantly affected \cite{Lindberg02}. It is, therefore, possible that the apo-state of SOD1 --- the subject of our studies ---  is implicated in misfolding and aggregation.

Most Cu, Zn superoxide dismutases, including human SOD1, have been shown to undergo irreversible aggregation on exposure to temperatures higher than their respective melting temperatures. The presence of {\em free} cysteines --- those that do not form disulfide bonds in the native state --- is known to be involved in this phenomenon. It is well-established that improper disulfide bond formation, concomitant with cysteine oxidation, possibly enhanced by metals in SOD1, is a cause of heat-denatured aggregation \cite{Perry87}. In the case of bovine SOD1, substitution by site-directed mutagenesis of free cysteine residues was found to greatly increase the reversibility of denaturation without substantially affecting the conformational stability \cite{McRee90, Lepock90}. Yeast SOD1 \cite{Lepock85} and {\em E. coli} SOD1 \cite{Battistoni98}, neither of which contain free cysteines, show partial and full reversibility respectively upon denaturation. However, SOD1 from {\em P. leiognathi} also denatures irreversibly despite having no free cysteines \cite{Bourne96}. Clearly, the presence of free cysteines is one of the crucial factors (but not the only factor) responsible for irreversible denaturation. One plausible reason for the effect of free cysteines is that the formation of improper disulfide bonds makes SOD1 more prone to kinetic traps. Therefore, we also study the effect of disulfide bonds on SOD1 folding reversibility.

An atomic resolution simulation of the folding process using traditional molecular mechanics force-fields is difficult  
by direct computational approaches because of the
vast dimensionality of the protein conformational space \cite{Karplus94a}.
Simplified models such as the G\={o} model
 \cite{go81, Abe81} provide a
powerful alternative to study folding because of their ability to
simulate folding on reasonable time scales and to
reproduce the basic thermodynamic and kinetic properties of 
proteins both on- \cite{Shimada01, Clementi00, Abkevich94} and off-lattices \cite{Dokholyan98b,
Dokholyan02b}. In the G\={o} model, the energy of the protein is expressed as a sum of pairwise native contact energies. A native contact exists between amino acids if they are closer to each other than a cut-off distance in the native state. Thus, in the G\={o} approximation, folding can be construed as a transition from a state with no or a small number of native contacts to a state with all native contacts.  
Despite their apparent simplicity, native topology-based approaches 
have yielded results in agreement with experiments, in particular for small proteins (\(<100\) amino
acids) \cite{Borreguero02, Ding02a} reflecting the underlying simplicity in the folding of these proteins. However, a G\={o} model of longer proteins may lead to discrepancy in protein folding dynamics between simulation and experiment because sequence-specific interactions may significantly alter the protein folding dynamics. Thus, we develop methodology to incorporate sequence specific information of the relative contribution of various amino-acids into the G\={o} model of SOD1.

\section{Results}

We perform DMD simulations using the unscaled G\={o} model, to determine the thermodynamic properties of a model
of SOD1 (as described in Methods). The temperature dependence of
average potential energy is shown in Fig.~1a. There is a  sigmoidal increase in the
potential energy with increasing temperature.  At low temperatures, SOD1
is present mostly in its native state, as evidenced by the small ($\leq 2.
83~$ \AA) root-mean-square deviation, $RMSD$, from the native state (Fig.~1b), and the small
fluctuations in the radius of gyration, \(R_{g}\), around its native state value (14.2 \AA) (data not shown).
 At high temperatures, SOD1 is unfolded with the $RMSD$ from the native state being greater than 40 \AA, demonstrating the loss of any structural similarity with the native state.
Also, $R_g$ is approximately three times its value in the native state (data not shown), indicating that the average distance between any two amino acids is approximately three times that in the native state.

In a typical trajectory (Fig.~2a) at the temperature corresponding to the mid-point of the abrupt change in potential energy (Fig.~1a) dependence on temperature (defined as the the folding
transition temperature $T_F$), we observe three distinct states: folded, unfolded and intermediate. A histogram of potential energies of
protein conformations near \(T_{F}\) is, therefore, tri-modal (Fig.~2c), instead of the 
expected bi-modal for a two-state protein, indicating the presence of
a folding intermediate. The detection of the intermediate state is in contrast with the experimental observation of the two-state folding dynamics of the SOD1 monomer \cite{Stroppolo00}. Thus, the {\em unscaled} G\={o} model for SOD1 does not accurately reproduce the folding dynamics of SOD1. 
Next, we calculate the pairwise interaction energies in the native state of SOD1 monomer and then employ the {\em scaled} G\={o} model, to uncover SOD1 folding thermodynamics and kinetics.

{\bf A few contacts make a large thermodynamic contribution to SOD1 native state stability.} We decompose the stability of SOD1 into pairwise contributions (as described in Methods) that are  plotted in
the lower triangle in Fig. 3 (the energy map).
In the energy map, some contacts, corresponding to both short-range and long-range interactions, have significantly stronger CHARMM interaction energies (from three to five times) than average.
We find that there are 16 long-range interactions that have large (about five times larger than the average) interaction energies:  Lys3-Glu21, Lys3-Glu153, Lys30-Glu100, Ser34-Asp96, Lys36-Asp92, Glu40-Lys91, Glu49-Arg115, Arg69-Glu77, His71-Asp83, Arg79-Asp83, Arg79-Asp101, His80-Asp83, Asp101-Val103, Glu121-Ser142, Asp125-Lys128, Glu133-Lys136.
The residues involved in the strong long-range contacts are highlighted in Fig.
4 and are involved in contacts between oppositely charged amino
acid pairs on the surface of the protein. Out of the 16
identified interactions, 6 have been found to be directly disrupted in patients
with FALS as a result of point mutations (listed in Table I) in the residues making these contacts \footnote{http://www.alsod.org}. In most cases, the mutations
correspond to the change of a charged amino acid for an
oppositely charged or a neutral one, suggesting an important role
these key contacts play in SOD1 folding.

{\bf Scaling thermodynamically important contacts modulates folding kinetics in the G\={o} model.} We test the importance of the identified amino acids in DMD
simulations by scaling their contact strength in the G\={o} potential. We increase the
depth of the interaction well of the identified key contacts to five times their original 
value in the {\em unscaled} G\={o} model ($b=5$).
Using this {\em scaled} G\={o} model,
we perform DMD simulations to calculate the thermodynamic properties of the SOD1. 
We find that the potential energy, $RMSD$ and $R_g$ (data not shown) exhibit the same sigmoidal trend as
for the {\em unscaled} G\={o} model, but the folding transition temperature increases to approximately 1.1
from its earlier value of 1.0. Remarkably, a typical
trajectory near the folding transition temperature ($T_F=1.15$) shows (Fig.~2b) that the intermediate state is no longer populated and the
protein only exists in either the folded or the unfolded state (Fig.~2b and d) near $T_F$.
 Thus, by strengthening these {\em key contacts} we reproduce the experimentally observed two state folding of SOD1 monomer \cite{Mei92, Stroppolo00}. As control, we randomly select several combinations of sixteen contacts and scale them by the same procedure. In such control simulations, we do not observe such effect on the folding kinetics (Fig.~2b), thus demonstrating that the effect of strengthening the identified contacts is specific. 
Importantly, this result uncovers a crucial connection between the folding thermodynamics and kinetics 
of SOD1: in our model, the residues that contribute the most to the SOD1 thermodynamic stability of the native state are also crucial for keeping the folding kinetics two-state.     

{\bf Improper disulfide bonds cause kinetic trapping of SOD1.} Human SOD1 is known to aggregate
irreversibly on heat denaturation \cite{Rodriguez02} {\em in vitro}. One of the reasons for the irreversibility is improper disulfide bond formation
between free cysteines. There
are four cysteine residues in the SOD1 monomer, at positions 6, 57, 111 and
146 of which residues 57 and 146 form a disulfide linkage in the native state. We study the effect of non-native disulfide bond formation
by starting from a fully unfolded conformation,
generated at T=2.3 (\(\gg T_{F}\)), and cooling the system to low
temperature T=0.4 (\(\ll T_{F}\)) slowly, while allowing the possibility of interactions
between any of the four cysteines to form improper disulfide
bond(s). We model disulfide bonds by an attractive potential ($b=2$, $\gamma=-1$ in Eq.~(2)). As a
control, we repeat the annealing procedure without these non-native
disulfide bonds ($b=1$, $\gamma=1$ in Eq.~(2)). We observe that out of 20 trajectories with improper
disulfide interactions, 4 result in SOD1 structures that are {\em trapped} in non-native conformations
while all of the 20 control trajectories fold to the correct native structure on the same time scale, implying that occurrence of improper disulfide linkage increases the number of kinetic traps for the SOD1 chain {\em en route} to its native state. The above observation is statistically significant (the probability of observing this result by chance: $p$-value $\approx$0.01). The increased number of kinetic traps makes intermediate states more populated and interaction of these intermediate states may make SOD1 prone to aggregation. We conclude that free cysteine residues play an important part in the kinetics of folding, and possibly, aggregation.

\section{Discussion}

Experimental evidence suggests that the unfolding of dimeric
superoxide dismutase occurs in a three state equilibrium \cite{Mei92, Stroppolo00}.

\begin{equation}
D{\rightleftharpoons}2M{\rightarrow}U 
\label{eq:sucks}
\end{equation}
where D, M and U are SOD1 dimer, monomer and unfolded state respectively. 
In order to understand the effect of mutations on SOD1 folding, it is 
important to study the effect of mutation on both processes in Eq.~(\ref{eq:sucks}).  Here, we study the second process (\(M{\rightarrow}U\)) 
and report thermodynamically and kinetically important residues
for SOD1 monomer folding.

Using our hybrid method, which incorporates molecular mechanics energetics into the G\={o} model, we identify the key residues involved in the
folding of the protein. Surprisingly, the key residues are found on the
surface of the protein and not in its hydrophobic core (Fig.~4). By comparing
the identified residues with a comprehensive list of mutations associated with FALS
\footnote{http://www.alsod.org}, we find that, 6 out of the 16 contacts
identified in our simulations are found to be implicated in the 
disease. In Table I, we list those FALS mutations, that
disrupt contacts identified from our simulations. We find that these mutations
replace a charged residue in the wild-type protein to a neutral or
oppositely charged residue in the mutant, thereby disrupting the electrostatic interaction. 
The presence of mutations disrupting these contacts further underscores the
importance of these key contacts and indicates that the disruption of
even one of these contacts may potentially result in protein
misfolding.

There have been no experimental studies characterizing SOD1 monomer folding kinetics. However, other members of the immunoglobulin-like (Ig-like) superfold family \cite{Murzin95}, to which SOD1 belongs, are experimentally well-characterized. Clarke and colleagues have established that the folding of pathways of Ig-like proteins share some common features \cite{Clarke99}. They observe that for five structurally similar Ig-like proteins, the folding rates are correlated with thermal stabilities, suggesting a similar folding pathway for all members of the family. Thus, for Ig-like proteins, the kinetic importance of residues in folding ($\phi$-values \cite{Fersht95}) is largely determined by the topology of the fold.  We, therefore, expect that the residues important for SOD1 folding are in structurally equivalent positions to other members of the Ig-like super-fold family, whose folding kinetics is experimentally well-studied \cite{Clarke99}. We choose Tnf3, the fibronectin type III domain of human tenascin, for which, based on $\phi$-values, eight residues in the hydrophobic core were identified as part of the folding nucleus \cite{Hamill00}. We identify residues in SOD1 that are structurally equivalent to the nucleus residues in Tnf3 by performing a structural alignment of the two proteins using DALI \cite{Holm95} and list these residues in Table II. We discover that two of the residues corresponding to this putative nucleus correspond to the strongly interacting key residues identified by our simulations, while three others are nearest neighbors of other residues we identify. This is significant because (i) mutations in the nucleus residues, being in proximity to the identified key residues, may disrupt the network of interactions of key residues and affect folding kinetics and (ii) in the G\={o} model entropic contributions are included as effective energetic interactions. Therefore, the identified key residues may be kinetically crucial because of the entropic contribution of the corresponding region of the structure (which is reflected by high $\phi$-values of the neighboring nucleus residues). Therefore,  kinetic evidence further supports our hypothesis that residues important for SOD1 native thermodynamics are important for folding kinetics.   

Richardson and Richardson have  pointed out \cite{Richardson02} that the interactions between
edges of \(\beta\)-strands are crucial to regulate protein aggregation propensity.
Naturally occurring \(\beta\)-sheet
proteins prevent aggregation by protecting the edges of
\(\beta\) strands through a multitude of fold-specific mechanisms. In
particular, for Ig-like \(\beta\)-sandwiches, 
strategically placed side-chain charges at the edges and
loop-crossing are dominant mechanisms which shield the edges of the
sheets. A recently solved crystal structure of mutant apo-SOD1 supramolecular assembly \cite{Elam03} identified three regions of the SOD1 structure where non-native gain of interaction was observed leading to amyloid like arrangement of SOD1 in crystals: 
the cleft between strands S5 and S6 (residues 83-100), 
the zinc loop (residues 65-78) and the electrostatic loop (residues 125-142). It was noted that disorder in the zinc and electrostatic loops promoted non-native interactions with the S5-S6 cleft, which is the interface between the two $\beta$-sheets of SOD1 and hence a likely site for edge-to-edge aggregation \cite{Richardson02}. We find that most of the key contacts for proper folding are made by residues in the three regions identified above. In particular, we see that contacts on both the edge-strands S5 and S6 play a crucial role in folding kinetics. The contacts made by residues Asp92, Asp96 and Lys91 which form the turn in the S5-S6 hairpin are crucial to seal the interface between the two $\beta$-sheets and likely prevent gain of non-native interactions, as shown in Fig.~5.  
Thus, our simulations constitute another piece in the growing body of evidence which implicates edge protection as the dominant mechanism by which natural proteins avoid aggregation. We postulate that the disruption of key interactions is a plausible
 scenario for mutant mediated
aggregation of SOD1. The re-organization of the S5-S6 edge may be the dominant feature which distinguishes the aggregation-prone misfolded state of SOD1, from the native state. The disruption of charge-charge interactions of
the side-chains of the identified key residues, which stabilize the $\beta$-barrel architecture of SOD1, might expose the edges of \(\beta\)-strand main-chain to interact with other structures, thus leading to
aggregation. Wang and Hecht \cite{Wang02} have shown that strategically designed mutations in \(\beta\)-strand
overhangs disfavor aggregation of normally
aggregating proteins, further supporting our hypothesis.

We show that the presence of oxidisable cysteines enhances the
kinetic trapping during folding. In the presence of metals, cysteine
oxidation is enhanced suggesting a plausible cause
for the misfolding of SOD1. However, it has recently been suggested \cite{Lindberg02} that  
misfolding of SOD1 likely takes place in the apo-enzyme state, when the metals have not yet been delivered to SOD1. In the reducing environment of the cell, cysteine-mediated misfolding/aggregation of the apoenzyme is, therefore, not a likely mechanism of SOD misfolding/aggregation. However, our observations do suggest a generic mechanism for SOD1 misfolding especially for mutations that do not affect the native state conformational stability \cite{Rodriguez02}. In the scenario similar to the one we have presented for free cysteines, other mutations may affect the folding kinetics of SOD1 by introducing non-native interactions along the folding pathway, leading to kinetic trapping without substantially affecting the native state stability.
      
\section{Methods}
{\bf The {\em unscaled} G\={o} model.} We use the Discrete Molecular Dynamics (DMD) algorithm \cite{Dokholyan98b, Zhou97, Smith97} to study the folding thermodynamics of SOD1. DMD has recently been used to study the folding kinetics \cite{Borreguero02, Ding02a, Zhou99a, Zhou99b} and aggregation of proteins \cite{Ding02b, Smith01}. We model the SOD1 chain by the beads-on-a-string model developed by Ding et al. \cite{Ding02a} with beads corresponding to all $C_{\alpha}$ and $C_{\beta}$ atoms and constraints between neighboring beads to mimic real protein flexibility. We study the folding of one monomer derived from the crystal structure of SOD1 (Protein data bank access code: 1SPD). We use the G\={o} potential \cite{go81} to model the interaction energy, $V_{ij}$, between $C_{\beta}$ atoms ($C_{\alpha}$ for Gly) $i$ and $j$ of SOD1:    
\begin{eqnarray}
  V_{ij} \, = \, \left \{
    \begin{array}{ll}
      +\infty,                      & \mbox{ $|r_i-r_j| \leq r_0$ }\\
      \gamma u_{ij},    & \mbox{ $r_0<|r_i-r_j| \leq r_1$ }\\
      0,                            & \mbox{ $|r_i-r_j| > r_1$ }
    \end{array}
  \right. \, ,
  \label{eq:potential}
\end{eqnarray}where $|r_i-r_j|$ is the distance between atoms $i$ and $j$. The parameters $r_0$ and  $r_1$ are the hard-core diameter (set to 3.25 \AA) and the cut-off distance (7.5 \AA) between $C_{\beta}$ atoms ($C_{\alpha}$ for glycine) in the native state respectively. We assign attractive interaction for native contacts ($\gamma= -1$), defined to exist between pairs of residues whose $C_{\beta}$ atoms ($C_{\alpha}$ for glycine) are closer than 7.5 \AA ~in the native state, and repulsive interaction ($\gamma= 1$) for non-native contacts, defined to exist between $C_{\beta}$ atoms ($C_{\alpha}$ for glycine) which are farther than 7.5 \AA ~in the native state.

The depth of the attractive square-well $u_{ij}$ in Eq.~(\ref{eq:potential}) determines the strength of the attractive interaction between two amino acids. First,
we consider the case where all contacts are equally strong, i.e. the
depth of each attractive well is $u$. We call this case the {\em unscaled} G\={o}
model and use it to perform DMD simulations of SOD1. We find that our model undergoes a collapse transition to its native state through a meta-stable intermediate (see Results) in contrast with the experimental observation of two-state folding dynamics of SOD1 monomer \cite{Mei92, Stroppolo00}. A possible reason for the existence of meta-stable intermediates is that in the unscaled G\={o} model, all native contacts have equal strength. While such an  approximation holds for smaller proteins (such as Src SH3 \cite{Borreguero02}) with relatively low number of contacts ($\sim162$), SOD1 monomer has 495 native contacts and the assumption of equal contact strength overestimates the population of intermediate species. Thus, in order to include sequence specificity of native interactions, we increase the strength of few specific native contacts
according to an all-atom model of SOD1 described below by making the depth of the
attractive well of some chosen native contacts $bu_{ij}$, where $b$ is the scale factor. We
call this the {\em scaled} G\={o} model. By scaling G\={o} potential according to an all atom model of the protein, we 
effectively include the sequence information.

{\bf The {\em scaled} G\={o} model.} To determine the relative strength of each individual native contact in SOD1, we use an effective energy function to decompose the total energy of the protein as a sum of pairwise terms \cite{Paci02}:
\begin{equation}
E = \sum_{i} \sum_{j>i} E_{ij}\, ,
\end{equation} where the summation is taken is over all pairs of residues of SOD1 and \(E_{ij}\) is the energy of interaction between residues $i$ and $j$. To determine the terms $E_{ij}$, we perform simulations with the molecular mechanics force field CHARMM \cite{Brooks83} to construct an all
atom model of the SOD1 monomer in the native state, starting from its crystal
structure (PDB accession code:1SPD). Recently, higher resolution crystal structures of apo-SOD1 have appeared in the literature~\cite{Strange03}, but our results are not qualitatively affected by starting with the higher resolution structure. We add hydrogen atoms and remove any steric
clashes occurring in the structure by energy minimization. The interactions which contribute the most to the relative contact interaction strength are the non-bonded interactions and the solvent-induced interaction screening \cite{Paci02}:
\begin{equation}
E_{ij} = E_{ij}^{non-bonded} + G_{ij}\, .
\end{equation}We compute \(E_{ij}^{non-bonded}\) as the sum of the Lennard-Jones and electrostatic interaction energies. The solvation free energy, $G_{ij}$, measures
the screening of the pairwise interaction due to polarization of the
solvent.

We calculate $G_{ij}$ using an implicit description of the solvent \cite{Sharp94}.
We approximate the solvent as a continuous dielectric medium with a
dielectric constant (\(\epsilon = 80\)) of water and the protein as a
continuous medium with dielectric constant (\(\epsilon =
1\)) of vacuum. 
We compute the solvation energy by solving the
Poisson-Boltzmann equation using the finite difference method \cite{Nicholls91}
available in CHARMM. Following Dominy {\em et al.}~\cite{Dominy02}, in order to decompose the solvation energy into pairwise
contributions, we eliminate all charges
in the protein except those corresponding to residue $i$ and
calculate its ``self'' solvation energy \(U_{i}\). Similarly, we compute $U_{j}$ and $U_{ij}$ when charges corresponding
to both $i$ and $j$ are present. Then, the solvent mediated screening of the interaction is:
\begin{equation} 
G_{ij} = U_{ij} - { [U_{i} + U_{j}] }\, .
\end{equation}
So, we calculate the effective pairwise energy
contribution of a contact in the folded state.

The $E_{ij}$ values that we thus obtain correspond to the folded state of the protein. In order to calculate the pairwise contribution of residues to the stability, we subtract the value of the contact energy in the unfolded state from the $E_{ij}$ values obtained above. The unfolded state is an ensemble of conformations. It has been noted that there is considerable native like structure in the unfolded state \cite{Zagrovic02,Shortle01, Elcock99}, meaning that the average environment of a contact is likely the same as in the folded state, but this contact is formed less frequently. Therefore, we model the unfolded state by simulating the {\em unscaled} G\={o} model near $T_{F}$ and estimating the frequency of native contact formation. The total contribution of a given contact to the stability can be written as:

\begin{equation}
\Delta E_{ij}=E_{ij}(1-p_{ij})
\end{equation}where $E_{ij}$ is the contact energy in the folded state and $p_{ij}$ is the probability of forming the contact in the unfolded state. The energy and contact maps are plotted in Fig.~3.
 
\section{Conclusions}

We present the folding thermodynamic and kinetic analysis of SOD1 using a hybrid molecular dynamics method. We show that our simple model qualitatively
reproduces the folding of SOD1 observed in
experiments. We find that interactions at the edges of SOD1 \(\beta\)
sheets and the zinc, electrostatic and cross-over loops
 are the key interactions that modulate the folding of SOD1
suggesting that mutants destabilizing these interactions may make
the enzyme more prone to aggregation.  In particular, we identify the S5-S6 cleft (residues 83-100) as a likely site for reorganization to yield an aggregation prone misfolded state, an observation consistent with recent experimental evidence \cite{Elam03} and the {\em negative design} paradigm \cite{Richardson02}. We also provide a plausible
explanation for the phenomenon of increased reversibility of folding
upon mutation of free cysteine residues. We find good agreement between the positions identified by us to be crucial for folding kinetics and those identified by experiments \cite{Hamill00} on other members of the super-fold family (Ig-like proteins) to which SOD1 belongs. Further, we find that mutations at six of the residue contacts we identify are implicated in FALS.

\section{Acknowledgements}
We thank S. V. Buldyrev for his help with DMD simulations, 
B. N. Dominy for help with CHARMM calculations and J. M. Borreguero, L. J. Hayward, B. Kuhlman and Z. Xu for helpful
discussions. We acknowledge the support of the UNC-CH Research Council Grant. SDK acknowledges the support of UNC Molecular and Cellular Biophysics Training Program and Scholars for Tomorrow Fellowship of UNC Graduate School.

\bibliographystyle{jmb}
\bibliography{ndbiblio}

\begin{thebibliography}{}

\bibitem{Cleveland01}
Cleveland, D.~W. \& Rothstein, J.~D. (2001{\em{}}).
\newblock From {Charcot to Lou Gehrig}: deciphering selective motor neuron
  death in {ALS}.
\newblock {\em Nature Rev. Neurosci.} {\bf 2}, 806--819.

\bibitem{Beckman01}
Beckman, J.~S., Estevez, A.~G., Crow, J.~P.  \& Barbeito, L. (2001{\em{}}).
\newblock Superoxide dismutase and the death of motorneurons in {ALS}.
\newblock {\em Trends Neurol. Sci.} {\bf 24}, S15--S20.

\bibitem{Rowland01}
Rowland, L.~P. \& Shneider, N.~A. (2001{\em{}}).
\newblock Amyotrophic lateral sclerosis.
\newblock {\em N. Engl. J. Med.} {\bf 344}, 1688--1700.

\bibitem{Siddique91}
Siddique, T. (1991{\em{}}).
\newblock Molecular genetics of familial amyotrophic lateral sclerosis.
\newblock {\em Adv. Neurol.} {\bf 56}, 227--231.

\bibitem{Rosen93}
Rost, D.~R., Siddique, T., Patterson, D., Figlewicz, D.~A., Sapp, P., Hentati,
  A., Donaldson, D., Goto, J., Oregan, J.~P., Deng, H.~X., Rahmani, Z., Krizus,
  A., McKennayasek, D., Cayabyab, A., Gaston, S.~M., Berger, R., Tanzi, R.~E.,
  Halperin, J.~J., Herzfeldt, B., Vandenbergh, R., Hung, W.~Y., Bird, T., Deng,
  G., Mulder, D.~W., Smyth, C., Laing, N.~G., Soriano, E., Pericakvance, M.~A.,
  Haines, J., Rouleau, G.~A., Gusella, J.~S., Horvitz, H.~R.  \& Brown, R.~H.
  (1993{\em{}}).
\newblock Mutations in {Cu/Zn} superoxide-dismutase gene are associated with
  familial amyotrophic-lateral-sclerosis.
\newblock {\em Nature, } {\bf 362}, 59--62.

\bibitem{Gurney94}
Gurney, M.~E., Pu, H., Chiu, A.~Y., Canto, M. C.~D., Polchow, C., Alexander,
  D.~D., Caliendo, J., Hentati, A., Kwon, Y.~W., Deng, H.~X., Chen, W., Zhai,
  P., Sufit, R.~L.  \& Siddique, T. (1994{\em{}}).
\newblock Motor neuron degeneration in mice that express a human {Cu,Zn}
  superoxide dismutase mutation.
\newblock {\em Science, } {\bf 264}, 1772--1775.

\bibitem{Gaudette00}
Gaudette, M., R.~Hirano, M.  \& Siddique, T. (2000{\em{}}).
\newblock Current status of {SOD1} mutations in familial amyotrophic lateral
  sclerosis.
\newblock {\em Amyotroph. Lateral Scler. Other Motor Neuron Disord.} {\bf 1},
  83--89.

\bibitem{Johnston00}
Johnston, J.~A., Dalton, M.~J., Gurney, M.~E.  \& Kopito, R.~R. (2000{\em{}}).
\newblock Formation of high molecular weight complexes of mutant
  {Cu,Zn}-superoxide dismutase in a mouse model for familial amyotrophic
  lateral sclerosis.
\newblock {\em Proc. Natl. Acad. Sci. U. S. A.} {\bf 97}, 12571--12576.

\bibitem{Bruijn98}
Bruijn, L.~I., Houseweart, M.~K., Kato, S., Anderson, K.~L., Anderson, S.~D.,
  Ohama, E., Reaume, A.~G., R.~W.~Scott, R.  \& Cleveland, D.~W. (1998{\em{}}).
\newblock Aggregation and motor neuron toxicity of an {ALS}-linked {SOD1}
  mutant independent from wild-type {SOD1}.
\newblock {\em Science, } {\bf 281}, 1851--1854.

\bibitem{Durham97}
Durham, H.~D., Roy, J., Dong, L.  \& Figlewicz, D.~A. (1997{\em{}}).
\newblock Aggregation of mutant {Cu/Zn} superoxide dismutase proteins in a
  culture model of {ALS}.
\newblock {\em J. Neuropathol. Exp. Neurol.} {\bf 56}, 523--530.

\bibitem{Reaume96}
Reaume, A.~G., Elliott, J.~L., Hoffman, E.~K., Kowall, N.~W., Ferrante, R.~J.,
  Siwek, D.~F., Wilcox, H.~M., Flood, D.~G., Beal, M.~F., Jr, R. H.~B., Scott,
  R.~W.  \& Snider, W.~D. (1996{\em{}}).
\newblock Motor neurons in {Cu/Zn} superoxide dismutase-deficient mice develop
  normally but exhibit enhanced cell death after axonal injury.
\newblock {\em Nature Genet.} {\bf 13}, 43--47.

\bibitem{Bromberg99}
Bromberg, M.~B. (1999{\em{}}).
\newblock Pathogenesis of amyotrophic lateral sclerosis: a critical review.
\newblock {\em Curr. Opin. Neurol.} {\bf 12}, 581--588.

\bibitem{Cleveland99}
Cleveland, D.~W. (1999{\em{}}).
\newblock From {Charcot to Lou Gehrig}: mechanisms of selective motor neuron
  death in {ALS}.
\newblock {\em Neuron, } {\bf 24}, 515--520.

\bibitem{Elliot01}
Elliot, J.~L. (2001{\em{}}).
\newblock Zinc and copper in the pathogenesis of amyotrophic lateral sclerosis.
\newblock {\em Prog. Neuro-Psychopharmacol. \& Biol. Psychiat.} {\bf 25},
  1169--1185.

\bibitem{Tran99}
Tran, P.~B. \& Miller, R.~J. (1999{\em{}}).
\newblock Aggregates in neurodegenerative disease: crowds and power?
\newblock {\em Trends Neurol. Sci.} {\bf 22}, 194--197.

\bibitem{Bruening01}
Bruening, W., Roy, J., Giasson, B., Figlewicz, D.~A., Mushynski, W.~E.  \&
  Durham, H.~D. (2001{\em{}}).
\newblock Up-regulation of protein chaperones preserves viability of cells
  expressing toxic {Cu/Zn}-superoxide dismutase mutants associated with
  amyotrophic lateral sclerosis.
\newblock {\em J. Neurochem.} {\bf 72}, 693--699.

\bibitem{Walsh02}
Walsh, D.~M., Klyubin, I., Fadeeva, J.~V., Rowan, M.~J.  \& Selkoe, D.~J.
  (2002{\em{}}).
\newblock Amyloid-beta oligomers: their production, toxicity and therapeutic
  inhibition.
\newblock {\em Biochem. Soc. Trans.} {\bf 30}, 552--557.

\bibitem{Chiti00}
Chiti, F., Taddei, N., Bucciantini, M., White, P.~M., Ramponi, G.  \& Dobson,
  C.~M. (2000{\em{}}).
\newblock Mutational analysis of the propensity for amyloid formation by a
  globular protein.
\newblock {\em EMBO J.} {\bf 19}, 1441--1449.

\bibitem{Canet02}
Canet, D., Last, A.~M., Tito, P., Sunde, M., Spencer, A., Archer, D.~B.,
  Redfield, C., Robinson, C.~V.  \& Dobson, C.~M. (2002{\em{}}).
\newblock Local cooperativity in the unfolding of an amyloidogenic variant of
  human lysozyme.
\newblock {\em Nature Struct. Biol.} {\bf 9}, 308--315.

\bibitem{Ding02b}
Ding, F., Dokholyan, N.~V., Buldyrev, S.~V., Stanley, H.~E.  \& Shakhnovich,
  E.~I. (2002{\em{}}).
\newblock Molecular dynamics simulation of {C-Src SH3} aggregation suggests a
  generic amyloidogenesis mechanism.
\newblock {\em J. Mol. Biol.} {\bf 324}, 851--857.

\bibitem{Chiti02a}
Chiti, F., Calamai, M., Taddei, N., Stefani, M., Ramponi, G.  \& Dobson, C.~M.
  (2002{\em{}}).
\newblock Studies of the aggregation of mutant proteins {\em in vitro} provide
  insights into the genetics of amyloid disease.
\newblock {\em Proc. Natl. Acad. Sci. U. S. A.} {\bf 6}, 1005--1009.

\bibitem{Chiti02b}
Chiti, F., Taddei, N., Baroni, F., Capanni, C., Stefani, M., Ramponi, G.  \&
  Dobson, C.~M. (2002{\em{}}).
\newblock Kinetic partitioning of protein folding and aggregation.
\newblock {\em Nature Struct. Biol.} {\bf 9}, 137--143.

\bibitem{Dobson99}
Dobson, C.~M. (1999{\em{}}).
\newblock Protein misfolding, evolution and disease.
\newblock {\em Trends Biochem. Sci.} {\bf 24}, 329--332.

\bibitem{Hallewell91}
Hallewell, R.~A., Imlay, K.~C., Lee, P., Fong, N.~M., Gallegos, C., Getzoff,
  E.~D., J.~A.~Tainer, J., Cabelli, D.~E., Tekampolson, P., G.~T.~Mullenbach,
  G.  \& Cousens, L. (1991{\em{}}).
\newblock Thermostabilization of recombinant human and bovine {Cu,Zn}
  superoxide dismutases by replacement of free cysteines.
\newblock {\em Biochem. Biophys. Res. Commun.} {\bf 181}, 474--480.

\bibitem{McRee90}
McRee, D.~E., Redford, S., Getzoff, E.~D., Lepock, J.~R., Hallewell, R.  \&
  Tainer, J. (1990{\em{}}).
\newblock Changes in crystallographic structure and thermostability of a
  {Cu,Zn} superoxide-dismutase mutant resulting from the removal of a buried
  cysteine.
\newblock {\em J. Biol. Chem.} {\bf 265}, 14234--14241.

\bibitem{Parge92}
Parge, H.~E., Hallewell, R.~A.  \& Tainer, J.~A. (1992{\em{}}).
\newblock Atomic structures of wild-type and thermostable mutant recombinant
  human {Cu,Zn} superoxide-dismutase.
\newblock {\em Proc. Natl. Acad. Sci. U. S. A.} {\bf 89}, 6109--6113.

\bibitem{Lepock90}
Lepock, J.~R., Frey, H.~E.  \& Hallewell, R.~A. (1990{\em{}}).
\newblock Contribution of conformational stability and reversibility of
  unfolding to the increased thermostability of human and bovine superoxide
  dismutase mutated at free cysteines.
\newblock {\em J. Biol. Chem.} {\bf 265}, 21612--21618.

\bibitem{Stroppolo00}
Stroppolo, M.~E., Malvezzi-Campeggi, F., Mei, G., Rosato, N.  \& Desideri, A.
  (2000{\em{}}).
\newblock Role of the tertiary and quaternary structures in the stability of
  dimeric copper,zinc superoxide dismutases.
\newblock {\em Arch. Biochem. Biophys.} {\bf 377}, 215--218.

\bibitem{Mei92}
Mei, G., Rosato, N., Silva, N., Rusch, R., Gratton, E., Savini, I.  \&
  Finazzi-Agro, A. (1992{\em{}}).
\newblock Denaturation of human copper-zinc superoxide dismutase by guanidine
  hydrochloride: a dynamic fluorescence study.
\newblock {\em Biochemistry, } {\bf 31}, 7224--7230.

\bibitem{Rodriguez02}
Rodriguez, J.~A., Valentine, J.~S., Eggers, D.~K., Roe, J.~A., Tiwari, A.,
  {Brown Jr.}, R.~H.  \& Hayward, L.~J. (2002{\em{}}).
\newblock Familial amyotrophic lateral sclerosis-associated mutations decrease
  the thermal stability of distinctly metallated species of human copper/zinc
  superoxide dismutase.
\newblock {\em J. Biol. Chem.} {\bf 277}, 15932--15937.

\bibitem{Hart98}
Hart, P.~J., Liu, H., Pellegrini, M., Nersissian, A.~M., Gralla, E.~B.,
  Valentine, J.~S.  \& Eisenberg, D. (1998{\em{}}).
\newblock Subunit asymmetry in the three-dimensional structure of a human
  {CuZnSOD} mutant found in familial amyotrophic lateral sclerosis.
\newblock {\em Protein Science, } {\bf 7}, 545--555.

\bibitem{Lindberg02}
Lindberg, M.~J., Tibell, L.  \& Oliveberg, M. (2002{\em{}}).
\newblock Common denominator of {Cu/Zn} superoxide dismutase mutants associated
  with amyotrophic lateral sclerosis: decreased stability of the apo state.
\newblock {\em Proc. Natl. Acad. Sci. U. S. A.} {\bf 99}, 16607--16612.

\bibitem{Perry87}
Perry, L.~J. \& Wetzel, R. (1987{\em{}}).
\newblock The role of cysteine oxidation in the thermal inactivation of {T4}
  lysozyme.
\newblock {\em Protein Eng.} {\bf 1}, 101--105.

\bibitem{Lepock85}
Lepock, J.~R., Arnold, L.~D., Torrie, B.~H., Andrews, B.  \& Kruuv, J.
  (1985{\em{}}).
\newblock Structural-analyses of various {Cu,Zn} superoxide dismuatses by
  differential scanning clorimetry and {Raman-spectroscopy}.
\newblock {\em Arch. Biochem. Biophys.} {\bf 241}, 243--251.

\bibitem{Battistoni98}
Battistoni, A., Folcarelli, S., Cervoni, L., Polizio, F., Dessideri, A.,
  Giartosio, A.  \& Rotilio, G. (1998{\em{}}).
\newblock Role of the dimeric structure in {Cu,Zn} superoxide dismutase.
\newblock {\em J. Biol. Chem.} {\bf 273}, 5655--5661.

\bibitem{Bourne96}
Bourne, Y., Redford, S.~M., Steinman, H.~M., Lepock, J.~R., Tainer, J.~A.  \&
  Getzoff, E.~D. (1996{\em{}}).
\newblock Novel dimeric interface and electrostatic recognition in bacterial
  {Cu,Zn} superoxide dismutase.
\newblock {\em Proc. Natl. Acad. Sci. U. S. A.} {\bf 93}, 12774--12779.

\bibitem{Karplus94a}
Karplus, M. \& Shakhnovich, E.~I. (1994{\em{}}).
\newblock Protein folding: theoretical studies of thermodynamics and dynamics.
\newblock In {\em Protein Folding}, (Creighton, T., ed.),. W. H. Freeman and
  Company New York.

\bibitem{go81}
G\={o}, N. \& Abe, H. (1981{\em{}}).
\newblock Noninteracting local-structure model of folding and unfolding
  transition in globular proteins. i. formulation.
\newblock {\em Biopolymers, } {\bf 20}, 991--1011.

\bibitem{Abe81}
Abe, H. \& G\={o}, N. (1981{\em{}}).
\newblock Noninteracting local-structure model of folding and unfolding
  transition in globular proteins. ii. application to two-dimensional lattice
  proteins.
\newblock {\em Biopolymers, } {\bf 20}, 1013--1031.

\bibitem{Shimada01}
Shimada, J., Kussell, E.~L.  \& Shakhnovich, E.~I. (2001{\em{}}).
\newblock The folding kinetics and thermodynamics of crambin using an all-atom
  monte-carlo simulation.
\newblock {\em J. Mol. Biol.} {\bf 308}, 79--95.

\bibitem{Clementi00}
Clementi, C., Nymeyer, H.  \& Onuchic, J.~N. (2000{\em{}}).
\newblock Topological and energetic factors: what determines the structural
  details of the transition state ensemble and ``en-route'' intermediates for
  protein folding? an investigation for small globular proteins.
\newblock {\em J. Mol. Biol.} {\bf 278}, 937--953.

\bibitem{Abkevich94}
Abkevich, V.~I., Gutin, A.~M.  \& Shakhnovich, E.~I. (1994{\em{}}).
\newblock Specific nucleus as the transition state for protein folding:
  evidence from the lattice model.
\newblock {\em Biochemistry, } {\bf 33}, 10026--10036.

\bibitem{Dokholyan98b}
Dokholyan, N.~V., Buldyrev, S.~V., Stanley, H.~E.  \& Shakhnovich, E.~I.
  (1998{\em{}}).
\newblock Molecular dynamics studies of folding of a protein-like model.
\newblock {\em Folding \& Design, } {\bf 3}, 577--587.

\bibitem{Dokholyan02b}
Dokholyan, N.~V., Li, L., Ding, F.  \& Shakhnovich, E.~I. (2002{\em{}}).
\newblock Topological determinants of protein folding.
\newblock {\em Proc. Natl. Acad. Sci. U. S. A.} {\bf 99}, 8637--8641.

\bibitem{Borreguero02}
Borreguero, J.~M., Dokholyan, N.~V., Buldyrev, S., Stanley, H.~E.  \&
  Shakhnovich, E.~I. (2002{\em{}}).
\newblock Thermodynamics and folding kinetics analysis of the {SH3} domain from
  discrete molecular dynamics.
\newblock {\em J. Mol. Biol.} {\bf 318}, 863--876.

\bibitem{Ding02a}
Ding, F., Dokholyan, N.~V., Buldyrev, S.~V., Stanley, H.~E.  \& Shakhnovich,
  E.~I. (2002{\em{}}).
\newblock Direct molecular dynamics observation of protein folding transition
  state ensemble.
\newblock {\em Biophys. J.} {\bf 83}.

\bibitem{Murzin95}
Murzin, A.~G., Brenner, S.~E., Hubbard, T.  \& Chothia, C. (1995{\em{}}).
\newblock {SCOP}: {A} structural classification of proteins database for the
  investigation of sequences and structures.
\newblock {\em J. Mol. Biol.} {\bf 247}, 536--540.

\bibitem{Clarke99}
Clarke, J., Cota, E., Fowler, S.~B.  \& Hamill, S.~J. (1999{\em{}}).
\newblock Folding studies of immunoglobulin-like beta-sandwich proteins suggest
  that they share a common folding pathway.
\newblock {\em Structure, } {\bf 7}, 1145--1153.

\bibitem{Fersht95}
Fersht, A.~R. (1995{\em{}}).
\newblock Characterizing transition states in protein folding: an essential
  step in the puzzle.
\newblock {\em Curr. Opinion Struc. Biol.} {\bf 5}, 79--84.

\bibitem{Hamill00}
Hamill, S.~J., Steward, A.  \& Clarke, J. (2000{\em{}}).
\newblock The folding of an immunoglobulin-like greek key protein is defined by
  a common-core nucleus and regions constrained by topology.
\newblock {\em J. Mol. Biol.} {\bf 297}, 165--178.

\bibitem{Holm95}
Holm, L. \& Sander, C. (1995{\em{}}).
\newblock {DALI}: a network tool for protein structure comparison.
\newblock {\em Trends Biochem. Sci.} {\bf 20}, 478--480.

\bibitem{Richardson02}
Richardson, J.~S. \& Richardson, D.~C. (2002{\em{}}).
\newblock Natural $\beta$-sheet proteins use negative design to avoid
  edge-to-edge aggregation.
\newblock {\em Proc. Natl. Acad. Sci. U. S. A.} {\bf 99}, 2754--2759.

\bibitem{Elam03}
Elam, J.~S., Taylor, A.~B., Strange, R.~W., Antonyuk, S., Doucette, P.~A.,
  Rodriguez, J.~A., Hasnain, S.~S., Hayward, L.~J., Valentine, J.~S., Yeates,
  T.~O.  \& Hart, P.~J. (2003{\em{}}).
\newblock Amyloid-like filaments and water-filled nanotubes formed by {SOD1}
  mutant proteins linked to familial {ALS}.
\newblock {\em Nature Struct. Biol.} {\bf 10}, 461--467.

\bibitem{Wang02}
Wang, W.~X. \& Hecht, M.~H. (2002{\em{}}).
\newblock Rationally designed mutations convert de novo amyloid-like fibrils
  into monomeric beta-sheet proteins.
\newblock {\em Proc. Natl. Acad. Sci. U. S. A.} {\bf 99}, 2760--2765.

\bibitem{Zhou97}
Zhou, Y., Karplus, M., Wichert, J.~M.  \& Hall, C.~K. (1997{\em{}}).
\newblock Equilibrium thermodynamics of homopolymers and clusters: molecular
  dynamics and monte carlo simulations of system with square-well interactions.
\newblock {\em J. Chem. Phys.} {\bf 107}, 10691--10708.

\bibitem{Smith97}
Smith, S.~W., Hall, C.~K.  \& Freeman, B.~D. (1997{\em{}}).
\newblock Molecular dynamics for polymeric fluids using discontinuous
  potentials.
\newblock {\em J. Comput. Phys.} {\bf 134}, 16--30.

\bibitem{Zhou99a}
Zhou, Y. \& Karplus, M. (1999{\em{}}).
\newblock Interpreting the folding kinetics of helical proteins.
\newblock {\em Nature, } {\bf 401}, 400--403.

\bibitem{Zhou99b}
Zhou, Y. \& Karplus, M. (1999{\em{}}).
\newblock Folding of a model three-helix bundle protein: a thermodynamic and
  kinetic analysis.
\newblock {\em J. Mol. Biol.} {\bf 293}, 917--951.

\bibitem{Smith01}
Smith, A.~V. \& Hall, C.~K. (2001{\em{}}).
\newblock Protein refolding versus protein aggregation: computer simulations on
  an intermediate-resolution protein model.
\newblock {\em J. Mol. Biol.} {\bf 312}, 187--202.

\bibitem{Paci02}
Paci, E., Vendruscolo, M.  \& Karplus, M. (2002{\em{}}).
\newblock Native and non-native interactions along protein folding and
  unfolding pathways.
\newblock {\em Proteins: Struc. Func. Genet.} {\bf 47}, 379--392.

\bibitem{Brooks83}
Brooks, B.~R., Bruccoleri, R.~E., Olafson, B.~D., States, D.~J., Swaminathan,
  S.  \& Karplus, M. (1983{\em{}}).
\newblock {CHARMM}: a program for macromolecular energy, minimization, and
  dynamics calculations.
\newblock {\em J. Comput. Chem.} {\bf 4}, 187--217.

\bibitem{Strange03}
Strange, R.~W., Antonyuk, S., Hough, M.~A., Doucette, P.~A., Rodriguez, J.~A.,
  Hart, P.~J., Hayward, L.~J., Valentine, J.~S.  \& Hasnain, S.~S.
  (2003{\em{}}).
\newblock The structure of holo and metal-deficient wild-type human {Cu, Zn}
  superoxide dismutase and its relevance to familial amyotrophic lateral
  sclerosis.
\newblock {\em J. Mol. Biol.} {\bf 328}, 877--891.

\bibitem{Sharp94}
Sharp, K.~A. (1998{\em{}}).
\newblock Electrostatic interactions in macromolecules.
\newblock {\em Curr. Opinion Struc. Biol.} {\bf 3}, R108--R111.

\bibitem{Nicholls91}
Nicholls, A. \& Honig, B. (1991{\em{}}).
\newblock A rapid finite-difference algorithm, utilizing successive over
  -relaxation to solve the {Poisson-Boltzmann} equation.
\newblock {\em J. Comput. Chem.} {\bf 12}, 435--445.

\bibitem{Dominy02}
Dominy, B.~N., Perl, D., Schmid, F.~X.  \& {Brooks III}, C.~L. (2002{\em{}}).
\newblock The effects of ionic strength on protein stability: the cold shock
  protein family.
\newblock {\em J. Mol. Biol.} {\bf 319}, 541--554.

\bibitem{Zagrovic02}
Zagrovic, B., Snow, C.~D., Khaliq, S., Shirts, M.~R.  \& Pande, V.~S.
  (2002{\em{}}).
\newblock Native-like mean structure in the unfolded ensemble of small
  proteins.
\newblock {\em J. Mol. Biol.} {\bf 323}, 153--164.

\bibitem{Shortle01}
Shortle, D. \& Ackerman, M.~S. (2001{\em{}}).
\newblock Persistence of native-like topology in a denatured protein in 8 {$M$}
  urea.
\newblock {\em Science, } {\bf 293}, 487--489.

\bibitem{Elcock99}
Elcock, A.~H. (1999{\em{}}).
\newblock Realistic modeling of the denatured states of proteins allows
  accurate calculations of the {pH} dependence of protein stability.
\newblock {\em J. Mol. Biol.} {\bf 294}, 1051--1062.

\end{thebibliography}

\pagebreak


\begin{table}
\caption{Mutations in the identified kinetically important residues also found in patients with FALS. In the first column, 6 of the 16 contacts that we identify are shown. The second column lists mutations that disrupt these contacts and are found in FALS patients.}
\begin{tabular}{|l|l|l|}
\hline
Contact& Mutation(s)\\ \hline Lys3-Glu21&Glu21Lys, Glu21Gly\\\hline
Glu49-Arg115&Glu49Lys\\\hline Asp96-Ser34&Asp96Asn\\\hline Lys30-Glu100&Glu100Lys, Glu100Gly\\\hline Arg79-Asp101&Asp101Gly, Asp101Asn\\\hline Glu49-Arg115& Arg115Gly\\ \hline
\end{tabular}
\label{t:1}
\end{table}

\clearpage

\begin{table}
\caption{Comparison of kinetically important residues. Kinetically important in SOD1 were determined by (i) structural comparison with the protein 1TEN \cite{Hamill00} using DALI \cite{Holm95} and (ii) by our simulations. Nucleus represents high $\phi$-value residues in 1TEN, equivalent positions refer to structurally equivalent positions determined by DALI alignment and key residues in the vicinity of equivalent positions identified by our simulations are shown in column 3. N/A refers to no equivalent residue found.}
\begin{tabular}{|l|l|l|}
\hline
Nucleus (1TEN) & Equivalent Position (1SPD) & Key Residue\\ \hline Ile20&Phe31&Lys30\\\hline Tyr36&His48&Glu49\\\hline Ile48&Asp83&Asp83\\\hline Leu50
&Asn86&N/A\\\hline Ile59&Asp101&Asp101\\\hline Thr66&His110&N/A\\\hline Tyr68&Ile112&N/A\\\hline Val70&Gly114&Arg115\\ \hline
\end{tabular}
\label{t:2}
\end{table}
\clearpage

\begin{figure}[ht!]
\centerline{ \vbox{ \hbox{\epsfxsize=12cm
\epsfbox{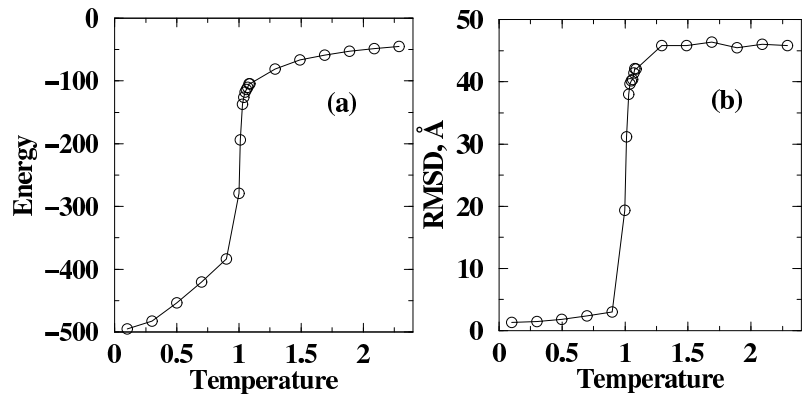}  }}}
\label{wall}
\caption{SOD1 thermodynamic properties.
Dependence on temperature of (a) average energy
(b) $RMSD$ in DMD simulations with {\em unscaled} G\={o} potential. The sigmoidal curves in both (a) and (b) show that SOD1 undergoes a collapse transition. Similar curves are obtained for DMD simulations with the {\em scaled} G\={o} potential.}
\end{figure}

\clearpage

\begin{figure}[ht!]
\centerline{ \vbox{ \hbox{\epsfxsize=10cm
\epsfbox{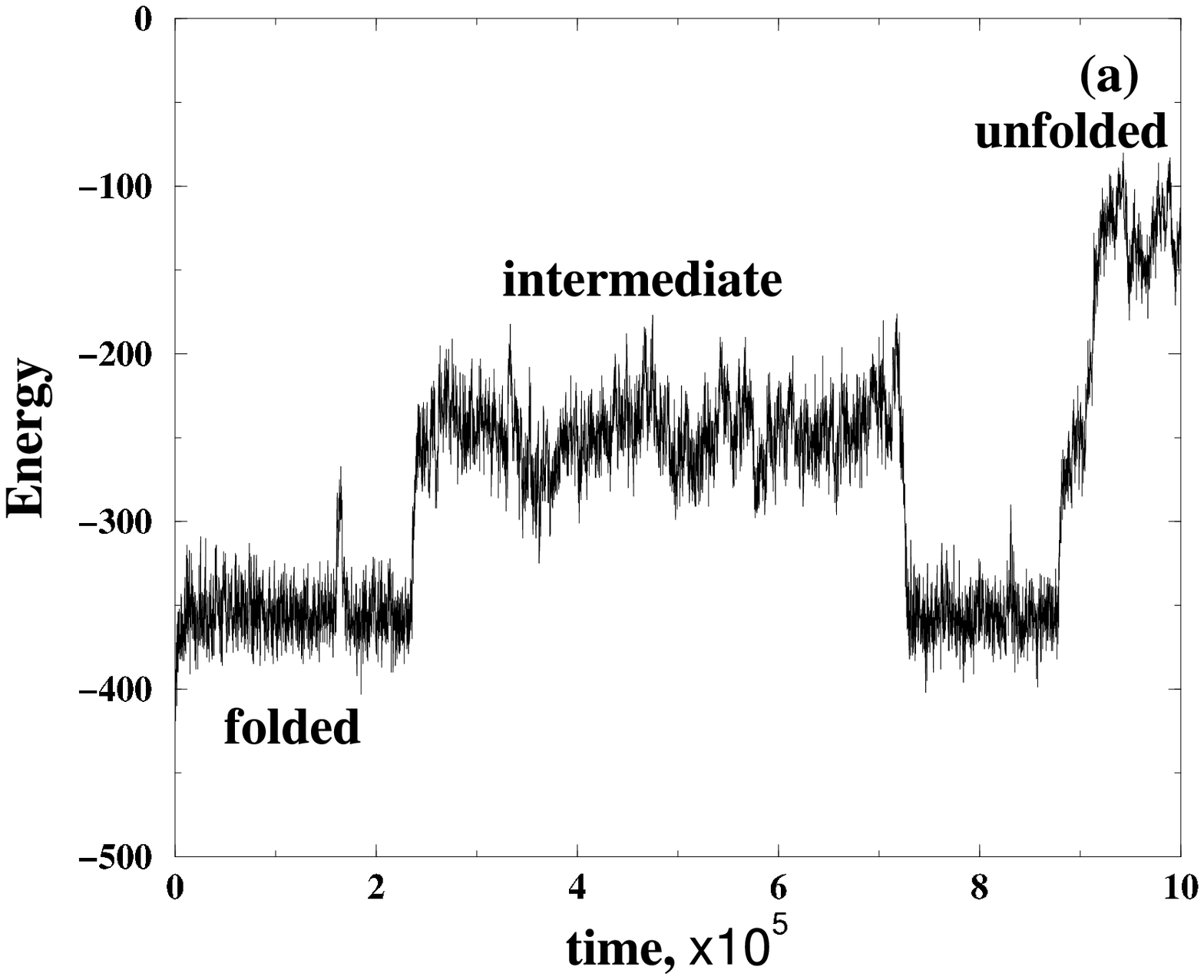}  }}
\vbox{ \hbox{\epsfxsize=10cm
\epsfbox{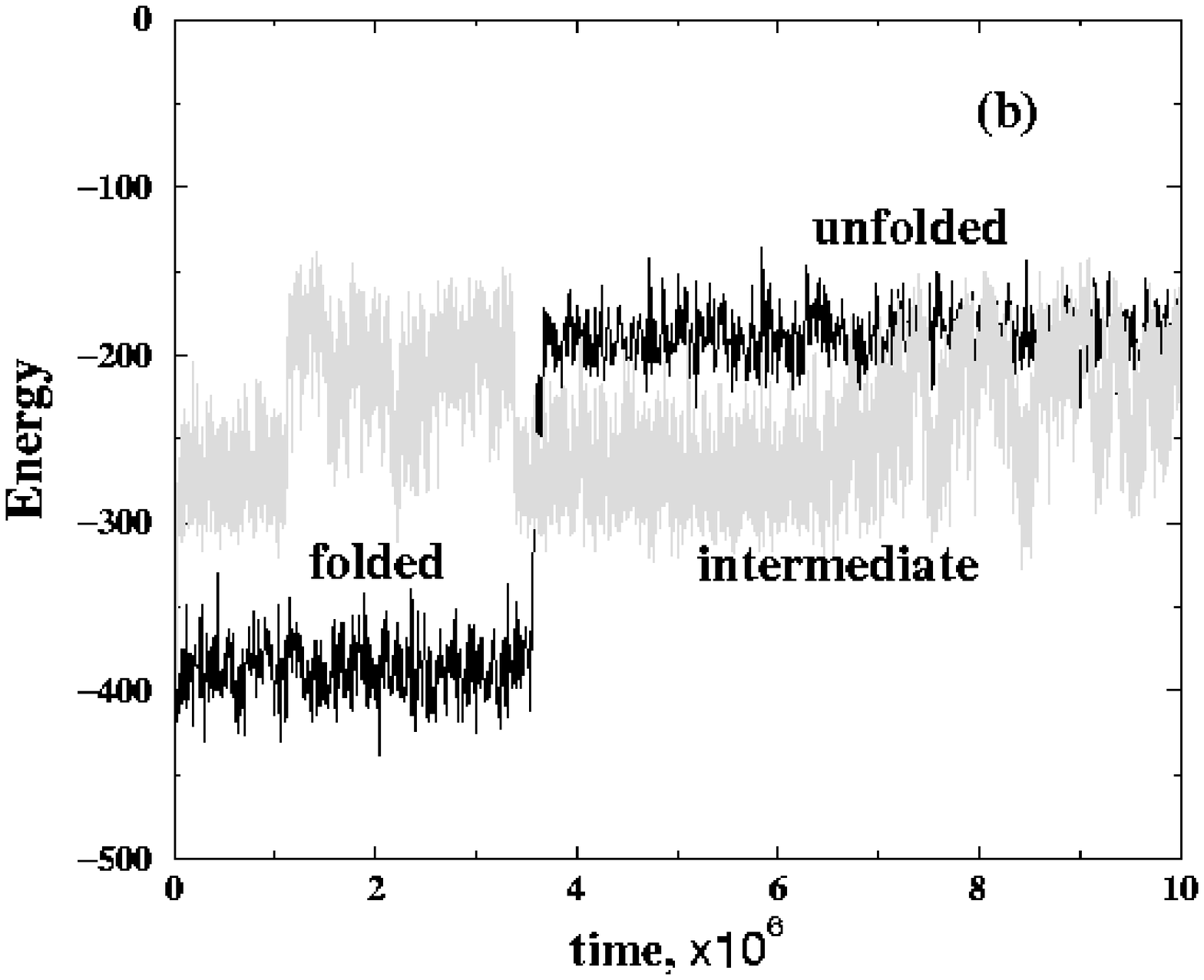}  }}}
\vspace*{0.5cm}
\centerline{ \vbox{ \hbox{\epsfxsize=10cm
\epsfbox{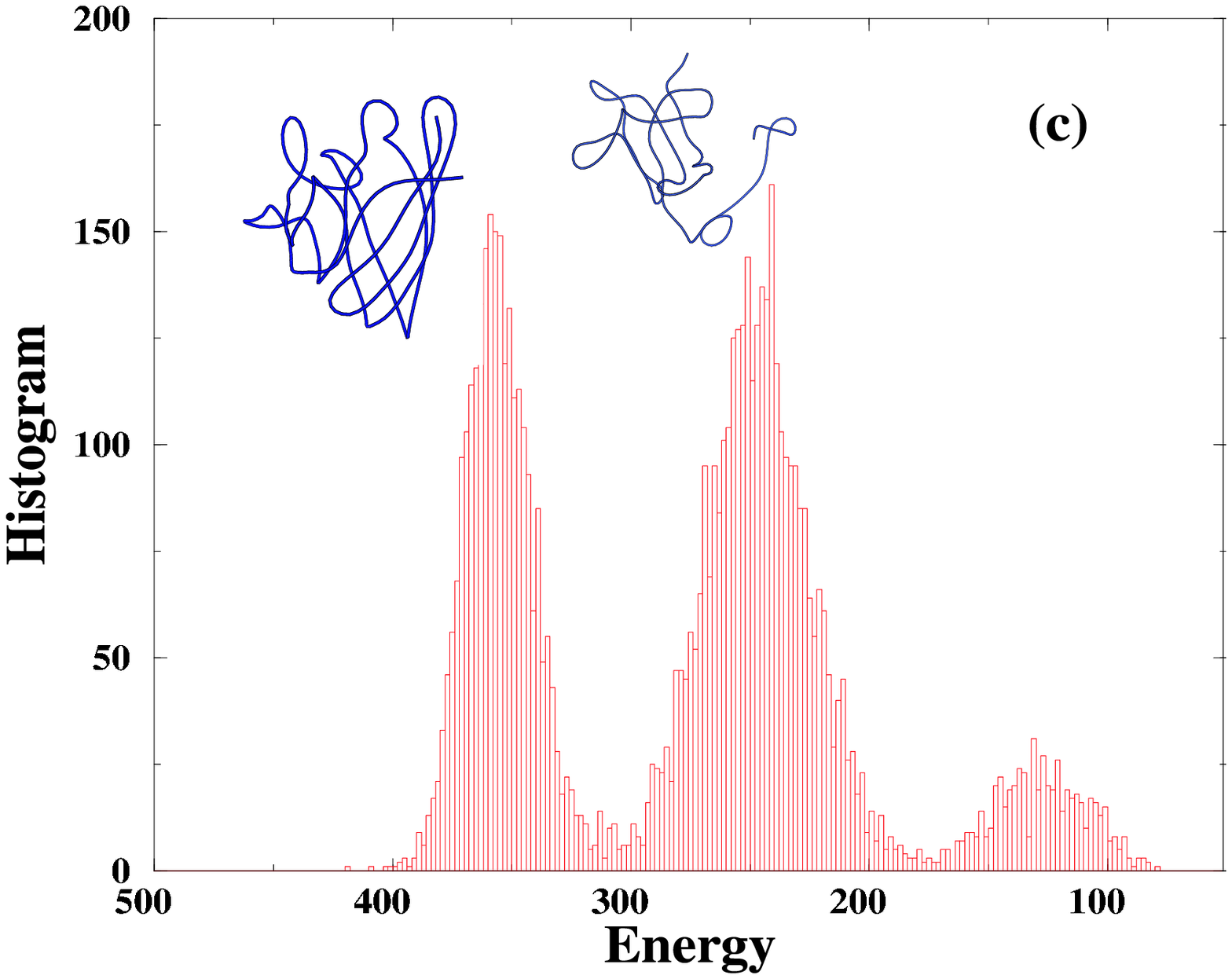}  }}
\vbox{ \hbox{\epsfxsize=10cm
\epsfbox{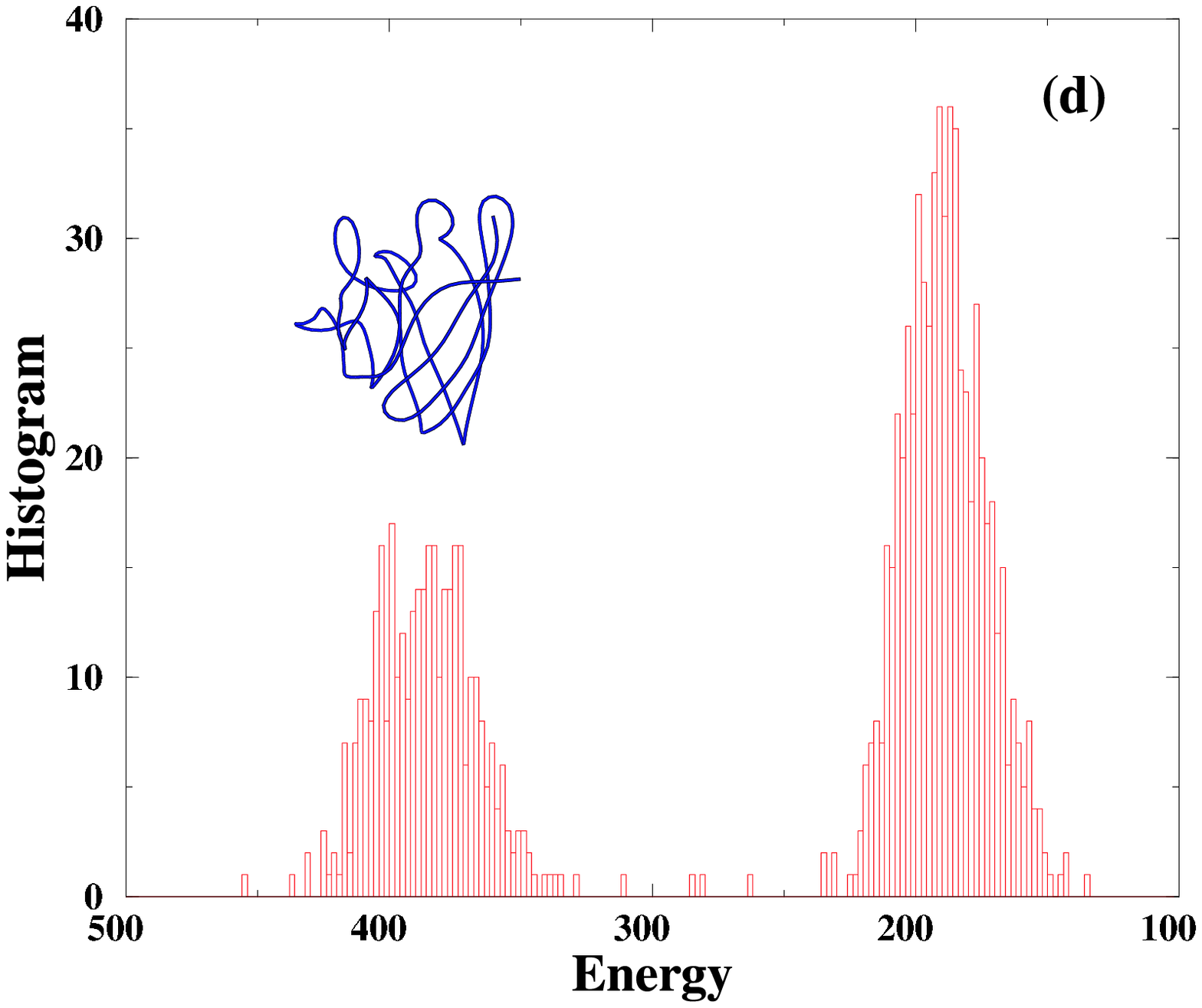}  }}}
\label{scale-unscaled}
\caption{Typical DMD trajectories near \(T_{F}\)
for (a) unscaled and (b) scaled G\={o} models. One trajectory is shown out of 4 for the unscaled G\={o} model and 5 for the scaled G\={o} model. In (b), a control trajectory with randomly chosen contacts (see text) strengthened is shown in grey. Histograms of energy values near $T_F$, for (c) unscaled and (d) scaled G\={o} models, averaged over 4 and 5 trajectories respectively, are also shown. From (b) and (d) we observe that the
 intermediate state is not populated in the scaled G\={o} model but is present in the unscaled G\={o} model, indicating that the kinetics of the scaled G\={o} model is indeed two-state as observed in experiments \cite{Mei92, Stroppolo00}.}
\end{figure}
\clearpage

\begin{figure}[ht!]
\centerline{ \vbox{ \hbox{\epsfxsize=13cm
\epsfbox{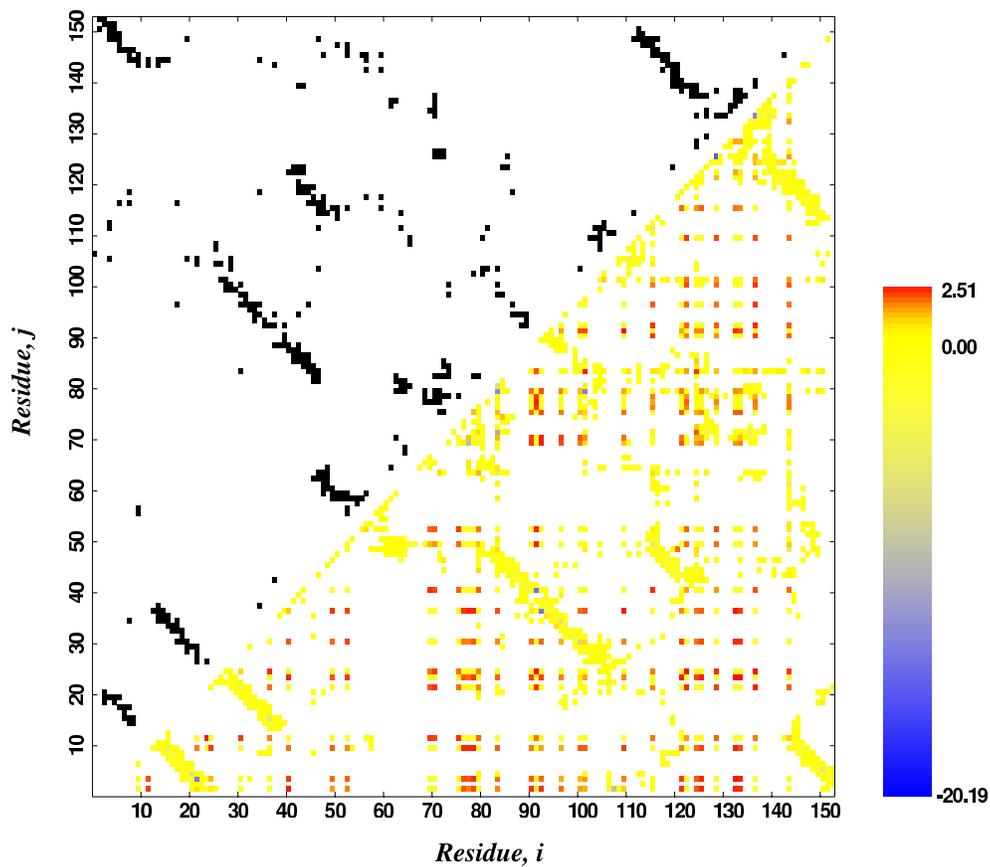}  }}}
\vspace*{0.5cm}
\label{contm}
\caption{Contact map (upper triangle) and CHARMM generated energy map (lower triangle) of the SOD1 monomer. The color shading in the energy map is assigned according to the interaction energy values; red indicates repulsive interactions and blue indicates attractive interactions. Energy values are in kcal/mol.}
\end{figure}
\clearpage

\begin{figure}[ht!]
\centerline{\vbox{ \hbox{\epsfxsize=10cm
\epsfbox{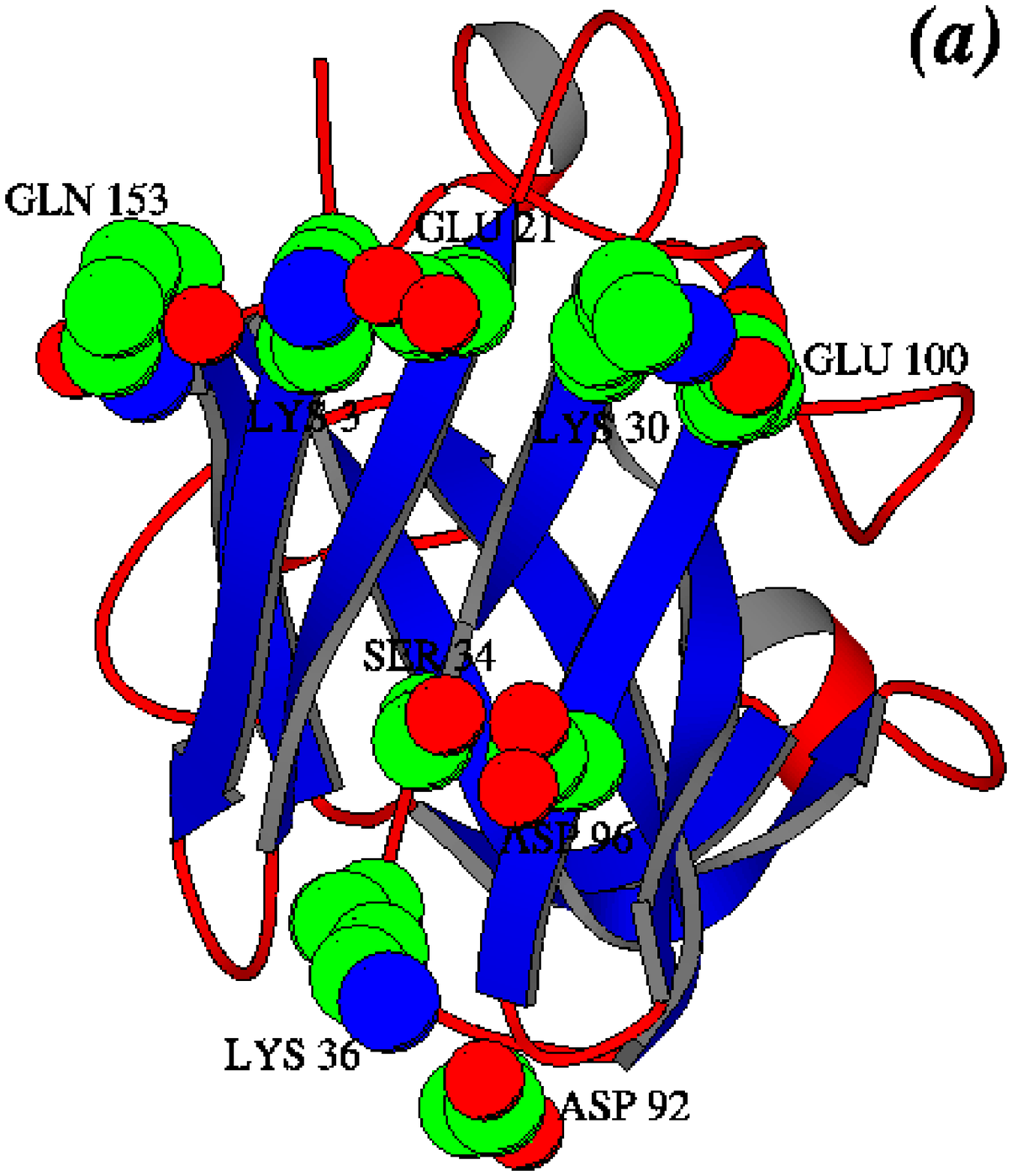} }}
\vbox{ \hbox{\epsfxsize=10cm
\epsfbox{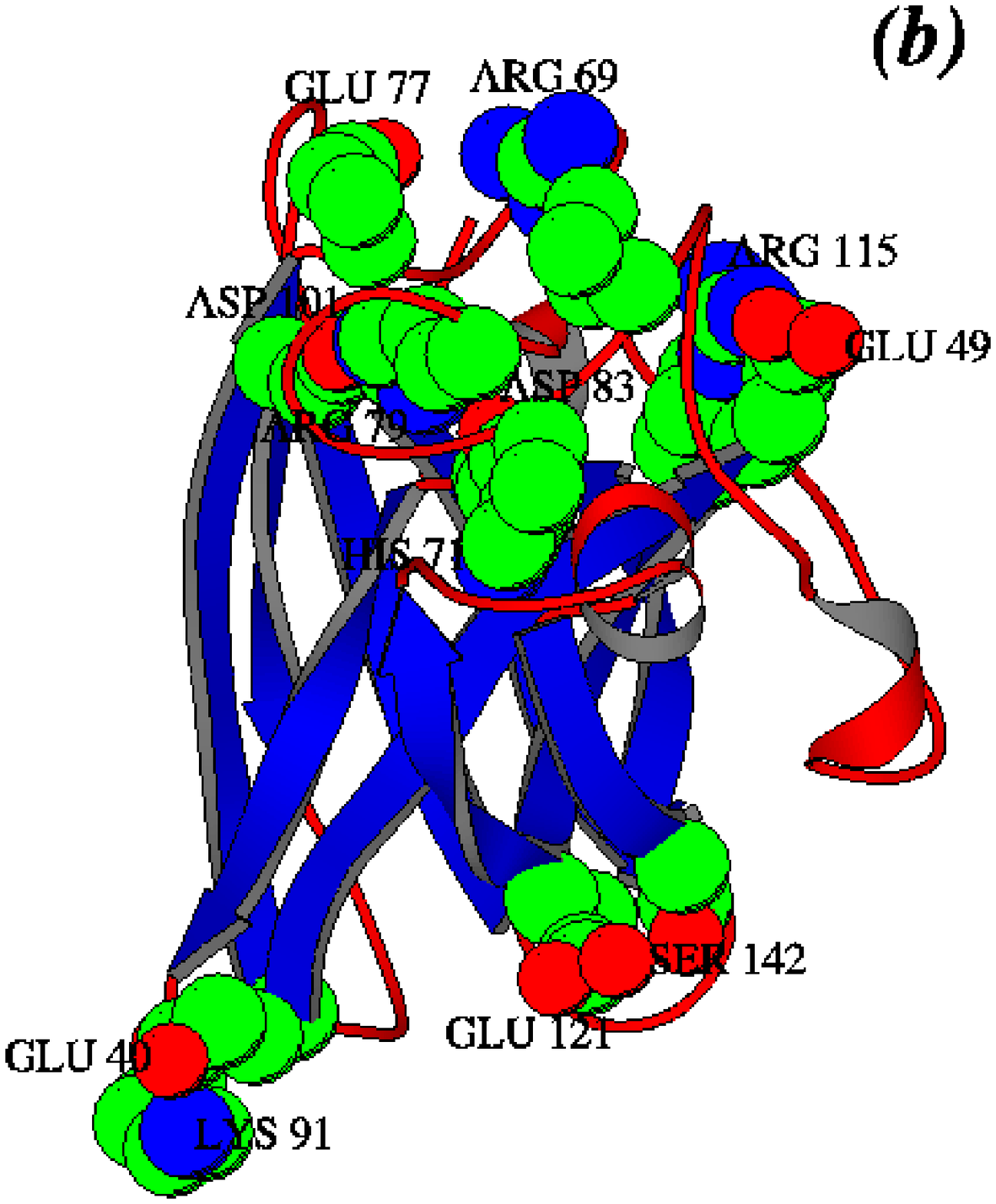} }}}
\vspace*{0.5cm}
\label{cvcomp}
\caption{Two projections (a) and (b) of SOD1 monomer structure highlighting residues making the identified contacts important for the two-state folding of SOD1: Lys3-Glu21, Lys3-Glu153, Lys30-Glu100, Ser34-Asp96, Lys36-Asp92, Glu40-Lys91, Glu49-Arg115, Arg69-Glu77, His71-Asp83, Arg79-Asp83, Arg79-Asp101, His80-Asp83, Asp101-Val103, Glu121-Ser142, Asp125-Lys128, Glu133-Lys136. Figure generated using Kraulis' Molscript program and co-ordinates from PDB structure 1SPD.}
\end{figure}
\clearpage

\begin{figure}[ht!]
\centerline{ \vbox{ \hbox{\epsfxsize=13cm
\epsfbox{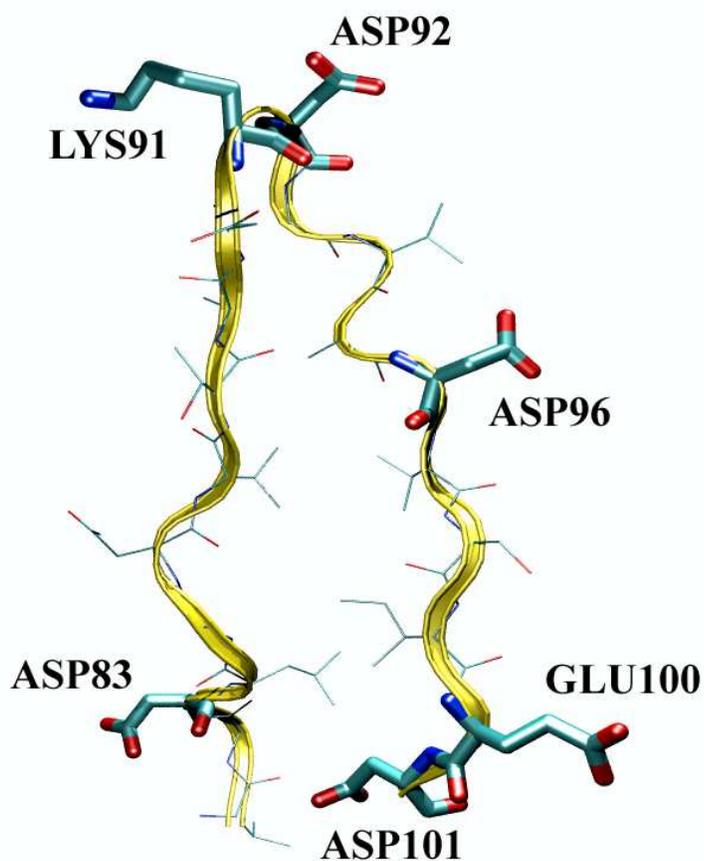}  }}}
\vspace*{0.5cm}
\label{cleft}
\caption{S5-S6 cleft. The residues in the cleft which are identified as crucial for making the folding kinetics two-state in our model, are highlighted. This cleft represents one of the edges of the $\beta$-barrel and is a possible site for initiation of non-native contact formation as also found in Ref. \cite{Elam03}, leading to aggregation.}
\end{figure}
\clearpage

\end{document}